\documentclass{elsart}

\date{30 October 1996}

\usepackage{amsmath,amsfonts,amssymb}   % for advanced math constructs
 
\usepackage[dvips]{epsfig}              % for inclusion of figures

\newcommand{\diff}[2][{}]{\frac{\partial#1}{\partial #2}}

\newcommand{\ddiff}[2][{}]{\frac{\partial^2 #1}{\partial {#2}^2}}

\newcommand{\mean}[1]{\langle{#1}\rangle}

\newcommand{\mcomma}{\;,}        
\newcommand{\mfullstop}{\;.}

\newcommand{\Pbed}[2]{\mathcal{P}\left(#1\,|\,#2\right)}
\newcommand{\Pf}[2]{\mathcal{P}_f\left(#1\,|\,#2\right)}
\newcommand{\Pre}[2]{\mathcal{P}_r\left(#1\,|\,#2\right)}

\DeclareMathOperator{\acot}{\cot^{-1}}

\DeclareMathOperator{\SNR}{SNR}

\newcommand{\myfig}[2]{
   \begin{figure}
      \begin{center}
         \setlength{\unitlength}{1mm}
         \begin{picture}(75,75)(0,0)
            \put( 0, 0){\epsfig{file=#1.eps,
                                 width=7.5cm,clip=}}
         \end{picture}
	\caption{#2}\label{#1}
      \end{center}
   \end{figure}
}

\begin{document}

\begin{frontmatter}

\title{Stochastic resonance in a model neuron with reset}

\author{H. E. Plesser\thanksref{corresp}\thanksref{hansmail}} \and
\author{S. Tanaka\thanksref{shigerumail}}
\address{Laboratory~for~Neural~Modeling, Frontier~Research~Program,
         RIKEN, 2-1~Hirosawa, Wako-shi, Saitama~351-01, Japan}

\thanks[corresp]{Corresponding author}
\thanks[hansmail]{E-mail: {plesser@prairie.riken.go.jp}}
\thanks[shigerumail]{E-mail: {shigeru@postman.riken.go.jp}}

\begin{abstract}
The response of a noisy integrate-and-fire neuron with reset to
periodic input is investigated.  We numerically obtain the
first-passage-time density of the pertaining Ornstein--Uhlenbeck
process and show how the power spectral density of the resulting spike
train can be determined via Fourier transform.  The neuron's output
clearly exhibits stochastic resonance.
\end{abstract}


\end{frontmatter}

%%%%%%%%%%%%%%%%%%%%%%%%%%%%%%%%%%%%%%%%%%%%%%%%%%%%%%%%%%%%%%%%%%%%

\section{Introduction}\label{sec:intro}
Neurons are inherently stochastic information processing devices,
whence the study of the influence of noise on neuronal signal
transmission and computation is of great interest.  Since the first
evidence for the enhancement of signals by noise was presented about 15
years ago~\cite{Benz:1981(L453)}, the phenomenon of stochastic
resonance has been demonstrated in a number of
physical~\cite{Fauv:1983(5),McNa:1988(2626)} and biological systems,
especially in sensory
neurons~\cite{Long:1991(656),Levi:1996(165),Wies:1995(33)}.  In the
wake of these experiments, the theory of stochastic resonance for
dynamic systems has been well
developed~\cite{McNa:1989(4854),Jung:1993(175)}, and was recently
extended to aperiodic signals~\cite{Coll:1995(R3321)}.

As neurons in higher centers of the brain need to maintain a high
signal-to-noise ratio as well as peripheral ones, it is plausible to
presume that stochastic resonance is a general principle of biological
information processing.  Indeed, models describing neurons as bistable
elements have been discussed in
detail~\cite{Zhou:1990(3161),Buls:1991(531)}.  For quantitative
comparison with neurophysiological data, though, model neurons closer
to biological reality need to be investigated.  To this end, we study
in this letter the response to a sinusoidal stimulus with superimposed
white noise of a widely used model neuron, the leaky integrate-and-fire
neuron which is reset upon firing~\cite{Tuck:Stoc}.  In this model, the
development over time of the membrane potential is given by the
solution of the Fokker--Planck equation describing the overdamped limit
of the Ornstein--Uhlenbeck process with an absorbing boundary.

Unfortunately, no analytic solution to this boundary value problem is
known~\cite{Lans:1995(397)}, while existing approximate solutions are
limited to particular parameter ranges; in particular, they require
sufficiently strong input noise~\cite{Buls:1996(3958)}.  Therefore, we
numerically solve for the first-passage-time density (FPTD), i.e.\ the
mathematical equivalent of the inter-spike-interval distribution (ISI),
using a computationally efficient integral equation approach.  From the
FPTD, we then calculate the power spectral density (PSD) of the spike
train generated by the model neuron via fast Fourier transform,
employing results from the theory of point processes. Finally, we
determine the signal-to-noise ratio (SNR) of the neuron's output, which
clearly exhibits stochastic resonance, i.e.\ SNR is maximal for a
finite strength of input noise.

Note the crucial difference between the model studied here and the
threshold detector model that has been studied by several authors in
recent years~\cite{Wies:1994(2125),Jung:1994(2513),Ging:1995(191)}.
The former is reset after each firing, whence individual threshold
crossings are uncorrelated and the entire spike train constitutes a
renewal process. The latter, to which we shall refer as
\emph{continuous-mode} model, does not include a reset mechanism,
but assigns one spike to each threshold crossing in positive
direction.  Thus, individual crossings are correlated and the membrane
voltage is governed by the same Fokker--Planck equation as our model,
but with natural boundaries at $\pm\infty$, permitting analytical
treatment.  Indeed, Jung~\cite{Jung:1995(93)} has given a theory of
stochastic resonance in continuous-mode threshold detectors based on
the periodic asymptotic solution of this Fokker--Planck problem.

%%%%%%%%%%%%%%%%%%%%%%%%%%%%%%%%%%%%%%%%%%%%%%%%%%%%%%%%%%%%%%%%%%%%

\section{The model}\label{sec:model}
The membrane voltage $x(t)$ of the model neuron is governed by the
Langevin equation of the overdamped Ornstein--Uhlenbeck
process~\cite{Tuck:Stoc,Lans:1995(457),Inou:1995(209)}
\begin{equation}
  \tau_m \dot{x}(t) = -x(t) + \mu + q \cos(\omega t +\varphi) + \xi(t)
	\mcomma \label{eq:oupdef}
\end{equation}
where we have set the resting potential to $x=0$. The membrane
time-constant $\tau_m$ and the drift term $\mu$ are positive constants,
while $q$, $\omega$ and $\varphi$ are arbitrary real constants, and
$\xi(t)$ is Gaussian white noise with zero mean and autocorrelation
$\mean{\xi(t)\xi(t')}=2D\delta(t-t')$.  The initial condition is
$x(0)=0$, and the neuron fires upon reaching the threshold voltage
$x(t)=x_{\mathrm{th}}$: both $x$ and the phase of the input stimulus
$\omega t + \varphi$ are reset to their values at $t=0$.

To normalize variable values, we scale time as $\bar{t} = {t}/{\tau_m}$
and voltage as $\bar{x}(\bar{t}) = {x(t)}/{x_{\mathrm{th}}}$, so that
time constant and threshold become~$1$, whence $\bar{\mu}
={\mu}/{x_{\mathrm{th}}}$, $\bar{q} = {q}/{x_{\mathrm{th}}}$,
$\bar{\omega} = \tau_m {\omega}$, $\bar{\varphi}=\varphi$ and $\bar{D}
= {D\tau_m}/{x_{\mathrm{th}}^2}$.  Thus we obtain the the dimensionless
equation
\begin{equation}
  \dot{x}(t) = -x(t) + \mu + q \cos(\omega t +\varphi) + \xi(t)
	\mcomma \label{eq:oup}
\end{equation}
where we have dropped the bars immediately for compactness of
notation. 

As mentioned above, each approach to the threshold is independent of
the past, because of the reset upon firing.  Therefore, assuming a
spike train of infinite duration, the firing process is a stationary
renewal process~\cite{Cox:Stat}. We will solve the FPT problem in the
next section before examining the spike train as a whole in
sections~\ref{sec:psd} and~\ref{sec:sr}.

%%%%%%%%%%%%%%%%%%%%%%%%%%%%%%%%%%%%%%%%%%%%%%%%%%%%%%%%%%%%%%%%%%%%

\section{First-Passage-Time Density}\label{sec:fptd}
In this section, we present an efficient numerical method for the
computation of the FPTD
\begin{equation}
  \rho(t)\d t = 
    \Pr\left\{\text{$x(t)=x_{\mathrm{th}}=1$ in $[t, t+\d t)$ if
	            $x(t=0)=0$}\right\} \mcomma
  \label{eq:defrho}
\end{equation} 
the theoretical counterpart of the ISI distribution.

The Fokker--Planck equation corresponding to the Langevin
equation~\eqref{eq:oup} is~\cite{Kamp:Stoc(1992)}
\begin{equation}
  \begin{split}
  \diff{t}\Pbed{x,t}{x_0,t_0} = 
	&-\diff{x}
           (-x+\mu+q\cos(\omega t +\varphi))\Pbed{x,t}{x_0,t_0}\\
        &+{D}\ddiff{x}\Pbed{x,t}{x_0,t_0} \mcomma
  \end{split}
  \label{eq:fpe}
\end{equation}
where $\Pbed{x,t}{x_0,t_0}$ is the probability density that the
voltage is $x$ at time $t$ if it was $x_0$ at time $t_0 < t$.  The
model is thus specified by the initial and boundary conditions
\mbox{$\Pre{x,t}{0,0} = \delta(x)$}, \mbox{$\Pre{-\infty,t}{0,0} = 0$}
and \mbox{$\Pre{1,t}{0,0} = 0$}, where the index $r$ indicates
restriction to $x \in (-\infty,1]$.  No analytic solution is known for
this boundary value problem and an approximation based on the method
of images is valid for a limited range of parameters
only~\cite{Buls:1996(3958)}.

Following Schr\"odinger~\cite{schr:1915(289)}, we thus construct an
integral equation equivalent to the above boundary value problem,
utilizing the solution $\Pf{x,t}{x_0,t_0}$ of~\eqref{eq:fpe} for the
unrestricted Ornstein--Uhlenbeck process on the entire real axis, i.e.\
with boundary conditions $\Pf{\pm\infty,t}{x_0,t_0}=0$.  The solution
is~\cite{Jung:1993(175)}
\begin{equation}
  \Pf{x,t}{x_0,t_0} = \frac{1}{\sqrt{2 \pi \sigma^2(t)}}
     \exp\left[-\frac{(x-\mean{x(t)})^2}{2 \sigma^2(t)}\right] \mcomma 
  \label{eq:unrest}
\end{equation}
where the mean and variance of $x(t)$ are 
(writing $\eta = \acot{\omega}$)
\begin{align}
  \begin{split}
    \mean{x(t)} &= 
      \mu + \frac{q}{\sqrt{1+\omega^2}}\sin(\omega t+\varphi+\eta) \\ 
     &\quad + \e^{-(t-t_0)}\Bigl[x_0-\mu- \frac{q}{\sqrt{1+\omega^2}}
	 \sin(\omega t_0 +\varphi+\eta)\Bigr]  \mcomma 
  \end{split}
  \label{eq:mean}
 \\
  \sigma^2(t)& = D \left( 1-\e^{-2(t-t_0)} \right) \mfullstop
  \label{eq:variance}
\end{align}
Then, the FPTD $\rho(t)$ is given by the Volterra integral
equation~\cite{Kamp:Stoc(1992)}
\begin{equation}
  \Pf{1,t}{0,0} = \int_{0}^{t}\!\d s\, \Pf{1,t}{1,s} \rho(s) \mfullstop
  \label{eq:inteq}
\end{equation}
Due to the sine terms in~\eqref{eq:mean}, the kernel $\Pf{1,t}{1,s}$
of the above equation cannot be rewritten as a function of $t-s$ alone
and a solution by Laplace transform is not possible.  A description of
the FPTD via its moments cannot be obtained either, since such methods
are based on the Laplace transform of the
kernel~\cite{Ricc:1988(43),Sieg:1951(617)}.

We thus solve for $\rho(t)$ using standard computational
techniques.  Since the kernel has an integrable square-root
singularity at $t=s$, we rewrite \eqref{eq:inteq} as
\begin{equation}
  \Pf{1,t}{0,0} = r(t)\rho(t) 
	+ \int_0^t\!\d s \, \Pf{1,t}{1,s} [\rho(s)-\rho(t)]
	\mcomma
  \label{eq:subsing}
\end{equation}
with $r(t) = \int_0^t \Pf{1,t}{1,s} \d s$.  This integral can be
evaluated numerically and, discretizing time as $t_j = j h$ with
stepsize $h>0$, we obtain the following algorithm for calculating the
FPTD~\cite{Pres:Nume(1992)}
\begin{equation}
  \rho_0 = 0\mcomma\quad
  \rho_m = \frac{h\sum_{j=1}^{m-1} K_{m,j}\rho_j-g_m}% 
     {\frac{h}{2}K_{m,0}+h\sum_{j=1}^{m-1} K_{m,j} -r_m} \mcomma\quad
  m = 1, 2, \ldots
  \label{eq:rhosol}
\end{equation}
where \mbox{$K_{m,j} = \Pf{1,m h}{1,j h}$}, \mbox{$g_m=\Pf{1,m
h}{0,0}$}, \mbox{$r_m=r(m h)$}. $\rho_0=\rho(0)$ follows from the
initial conditions.

The algorithm defined by~\eqref{eq:rhosol} has proven to be stable and
reliable.  Over a wide range of parameter values, the calculated FPTDs
$\rho_m$ are strictly non-negative (if numerical noise of the order of
machine accuracy is excluded) and the norm of the distributions
approaches $1$ from below as the range of calculation is extended
towards larger $t$.

We found a different integral-equation approach 
\cite{Buon:1987(784),Gior:1989(20)} to be slightly less stable for
some interesting parameter values.  In regions where both algorithms
are stable, results agree well.  

%%%%%%%%%%%%%%%%%%%%%%%%%%%%%%%%%%%%%%%%%%%%%%%%%%%%%%%%%%%%%%%%%%%%

\section{Power Spectral Density}\label{sec:psd}
To calculate the power spectral density (PSD) of the neuron's output,
let us first consider a train of $M\:\delta$-spikes with 
inter-spike-intervals $\tau_j$ distributed according to the FPTD
$\rho(\tau_j)$: 
\begin{equation}
  f_M(t) = \sum_{m=1}^{M} \delta(t-t_m) \mcomma\quad
  t_m = \sum_{j=1}^m \tau_j \mcomma\quad
  t_1=0 \mfullstop
  \label{eq:traindef}
\end{equation}
Neglecting the exact shape of the spikes amounts merely to dropping a
form factor from the spectrum, while all statistically relevant
information is contained in the firing times $t_m$ (see
also~\cite{Jung:1995(93)}).

For $M\rightarrow\infty$, this process is a stationary renewal
process, the spectra of which have been extensively discussed in
mathematical literature~\cite{Bart:1963(264),Cox:Stat}, where they are
known as {\em Bartlett spectra}~\cite{Dale:Intr}.  Applications to
neuronal systems have been rare to our
knowledge~\cite{Perk:1967(391)}.

The one-sided power spectral density of the spike train $f_M(t)$ is
given by
\begin{gather}
  S_M(\varOmega) = 
    \overline{\tilde{f}_M(-\varOmega)}\tilde{f}_M(-\varOmega) 
      + \overline{\tilde{f}_M(\varOmega)}\tilde{f}_M(\varOmega) 
    = 2 \overline{\tilde{f}_M(\varOmega)}\tilde{f}_M(\varOmega)
  \label{eq:smdef} \\
\intertext{where the bar indicates complex conjugation and}
  \tilde{f}_M(\varOmega) =  
     \frac{1}{\sqrt{2 \pi t_M}}
       \int_{0}^{t_M}\!\d t\, f_M(t) \e^{-i\varOmega t}
  \label{eq:ffdef}
\end{gather}
is the Fourier transform.  Inserting \eqref{eq:traindef} and
\eqref{eq:ffdef} into \eqref{eq:smdef} yields~\cite{Cox:Stat}
\begin{gather}
  \begin{split}
  S_M(\varOmega) 
    & = \frac{1}{\pi t_M} \sum_{m,k=1}^M \e^{-i\varOmega(t_m-t_k)}\\
    & = \frac{1}{\pi}\frac{M}{t_M}
	\left\{1
          +\int_0^{\infty}\!\d t\, {h}_M(t) \e^{-i\varOmega t} 
          +\int_0^{\infty}\!\d t\, {h}_M(t) \e^{ i\varOmega t} 
        \right\}
   \end{split} \label{eq:smhm} \\
  \intertext{where we have defined}
  {h}_M(t) = \frac{1}{M} \sum_{j,k=1}^{M-1} \delta(t_{j+k}-t_j-t)
  \mfullstop
  \label{eq:hmdef}
\end{gather}
Integrating ${h}_M(t)$ over non-overlapping intervals would give
the autocorrelation histogram of the neuron firing times.
In the limit of an infinite spike train, we obtain
\begin{equation}
  {h}_M(t) \rightarrow h(t) 
  \quad \text{and} \quad
  t_M/M \rightarrow \mean{\tau} \quad
  (M\rightarrow\infty)\mcomma 
  \label{eq:inflimit}
\end{equation}
where $h(t)$ is the renewal density and $\mean{\tau}$ the mean
first-passage-time. Note that the renewal density $h(t)$ is
\emph{not} a probability density, but $h(t)\d t$ is the probability
for a spike to occur in $[t, t+\d t)$.

From the theory of renewal
processes~\cite{Cox:Stat} we have for $\varOmega \ne 0$ 
\begin{equation}
  \int_0^{\infty}\!\d t\, h(t) \e^{i\varOmega t} 
    = \frac{\tilde{\rho}(\varOmega)}{1-\tilde{\rho}(\varOmega)} 
  \mfullstop
  \label{eq:hfrho}
\end{equation}
Here, $\tilde{\rho}(\varOmega)$ is the Fourier transform of the FPTD
$\rho(\tau)$.
Performing the limit in \eqref{eq:smhm} and inserting
\eqref{eq:hfrho}, we obtain the one-sided PSD of the infinite spike
train 
\begin{equation}
 S(\varOmega) = 
  \frac{1}{\pi\mean{\tau}} \left\{ 1 +
    \frac{\tilde{\rho}(\varOmega)}{1-\tilde{\rho}(\varOmega)} 
	+ \frac{\tilde{\rho}(-\varOmega)}{1-\tilde{\rho}(-\varOmega)}
  \right\} \quad(\varOmega > 0).
 \label{eq:S}
\end{equation}
Using this result, we can compute the PSD directly from the FPTD by
means of a discrete Fourier transform.

\myfig{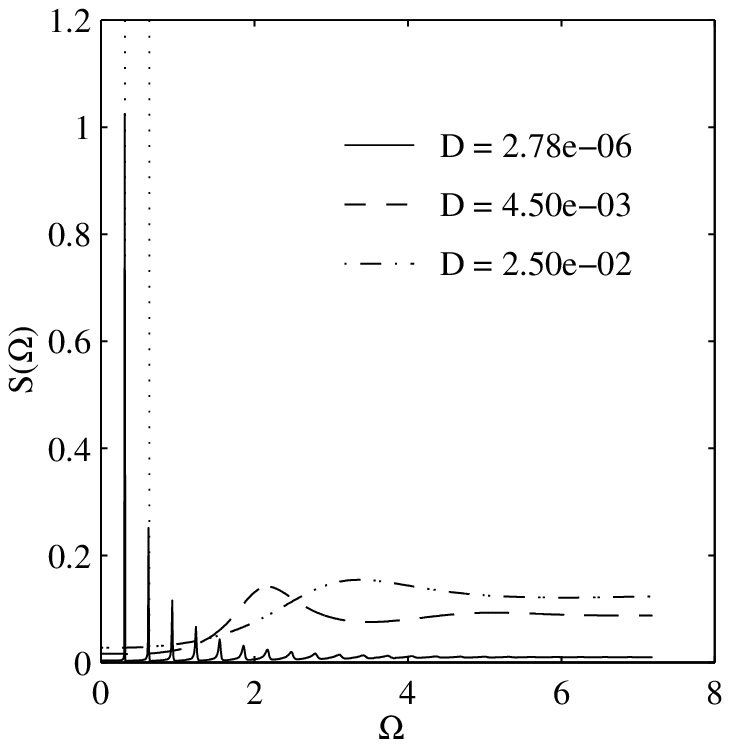}{Power spectral density for $\mu=0.97$, $q=0.03$,
$\omega=0.1\pi$, $\epsilon=0.0014$ for three different noise levels,
corresponding to $D_{\mathrm{max}}$, medium and high noise. The
vertical dotted lines mark the input frequency $\omega$ and its first
harmonic.} 

For white shot noise, i.e. the Poisson process with FPTD
$\rho(\tau)=\lambda\exp(-\lambda\tau)$, the terms in
$\tilde{\rho}(\varOmega)$ in \eqref{eq:S} cancel and a white spectrum
$S_P=1/\pi\mean{\tau}$ results.  Any deviation of $S(\varOmega)$ from
$S_P$ indicates the presence of a signal.  For the Ornstein--Uhlenbeck
process studied here, the spectra approach $S_P$ quickly for large
$\varOmega$ (Fig.~\ref{fig_spectra}).  We will therefore employ $S_P$
as the reference noise level in section~\ref{sec:sr}.

%%%%%%%%%%%%%%%%%%%%%%%%%%%%%%%%%%%%%%%%%%%%%%%%%%%%%%%%%%%%%%%%%%%%

\section{Stochastic resonance}\label{sec:sr}
Having set the mathematical stage, we may now explore the response of
our model neuron to sinusoidal input.  A single parameter
characterizing the input signal is the \emph{distance-from-threshold}
of the deterministic trajectory $\mean{x(t)}$
\begin{equation}
  \epsilon = 1 - \sup_{t\ge 0} \, \mean{x(t)} 
	= 1 - \Bigl(\mu+\frac{q}{\sqrt{1+\omega^2}}\Bigr) \mfullstop
  \label{eq:diffth}
\end{equation}
In defining the signal-to-noise ratio, the following difficulty
arises. The reset mechanism introduces a second timescale into the
system besides the one given by the input frequency.  Therefore, the
output spectrum instead of spikes will have maxima of finite width,
and the locations $\varOmega_{\mathrm{s}}$ of these are shifted away
from the input frequency $\omega$ (Fig.~\ref{fig_spectra}).  We thus
search a neighborhood of the input frequency for the signal peak and
define
\begin{equation}
\begin{split}
  \SNR & = \frac{\max\left\{S(\varOmega) | 
              \,(1-\alpha)\,\omega < \varOmega 
              < (1+\alpha)\,\omega  \right\}}{S_P} \\
       & = \pi\mean{\tau} \, {\max \left\{S(\varOmega) | 
	      \,(1-\alpha)\,\omega < \varOmega 
              < (1+\alpha)\,\omega  \right\}}
  \mfullstop
  \label{eq:snrdef}
\end{split}
\end{equation}
As discussed above, we use the uniform spectral density $S_P$ of the
Poissonian spike train with firing rate $1/\mean{\tau}$ as noise
reference level.  Note that no SNR is calculated if the spectrum is
monotonous in $\left[(1-\alpha)\,\omega,(1+\alpha)\,\omega \right]$.

For all data shown, we have calculated the FPTD $\rho(t)$ up to
\mbox{$t = t_{\mathrm{max}}$} such that
\mbox{$\int_0^{t_{\mathrm{max}}} \rho(t) \d t \ge 0.99$}.  Unless
stated otherwise below, we employed a stepsize of $h = 0.1$ and set the
initial phase of the stimulus to $\varphi=0$.  Parameter sets for which
$\rho(t)$ assumed negative values were discarded unless the latter
could clearly be identified as numerical noise.  PSDs were calculated
at increasing frequency resolutions until results became consistent.
The interval width for searching the signal was chosen as
$\alpha=0.07$.

\myfig{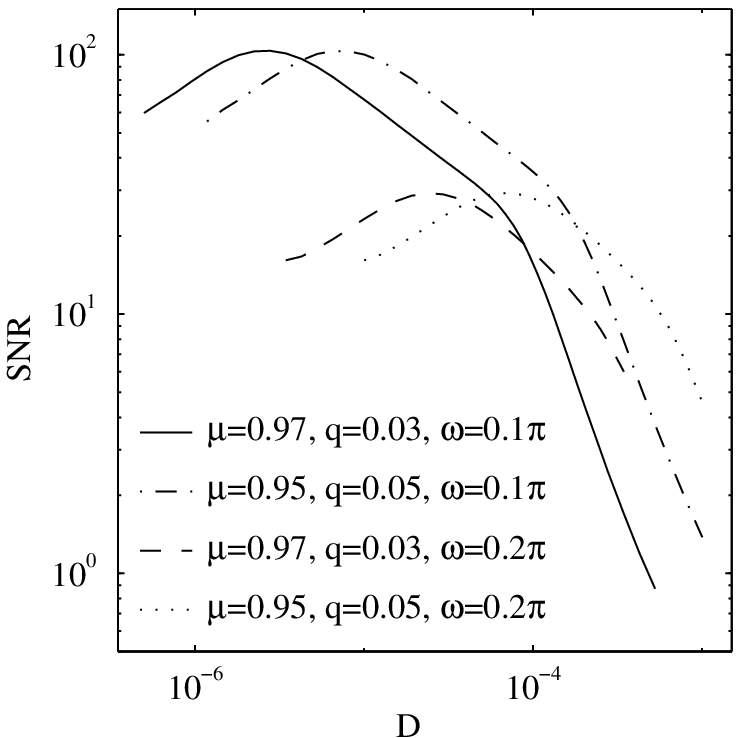}{Signal-to-noise ratio vs.\ input noise strength 
for signals with small distance-to-threshold.  
From top to bottom in the legend: $\epsilon=0.0014$,
$0.0023$,  $0.0046$, $0.0077$.}

\myfig{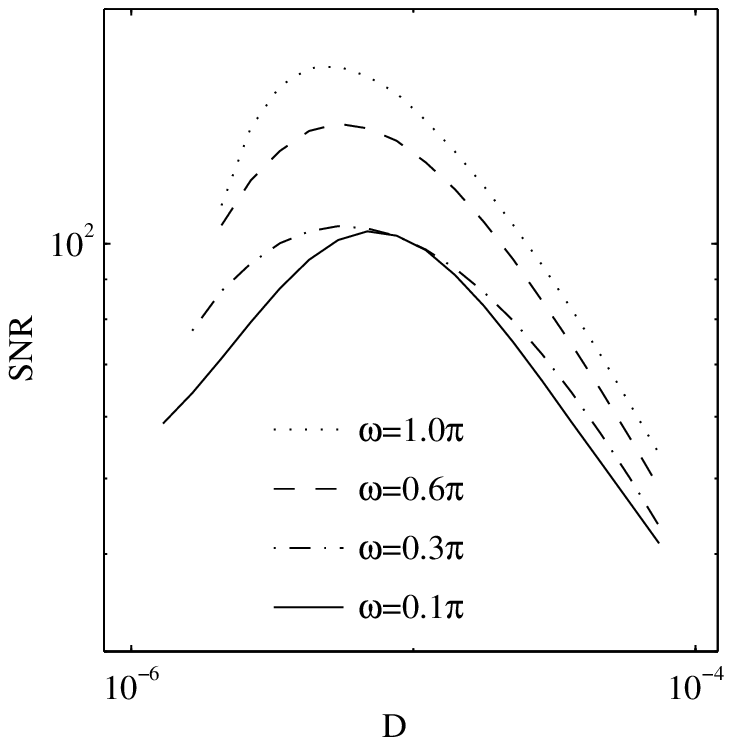}{Signal-to-noise ratio vs.\ input noise
strength for different input frequencies but the same
distance-from-threshold $\epsilon=0.0014$.  Here, $\mu=0.95$ and
$q=0.05\times\sqrt{1+\omega^2}/\sqrt{1+\omega_1^2}$, 
$\varphi=\acot\omega_1-\acot\omega$, $h=0.1\times\omega_1/\omega$,
$\omega_1=0.1\pi$.}

As the central result of our work, we show in
Figs.~\ref{fig_epsvar} and~\ref{fig_epsfix}
the dependence of the signal-to-noise ratio on the input noise
strength for various values of drift term $\mu$, modulation
amplitude $q$ and frequency $\omega$, which correspond 
to distances-from-threshold $0.001 < \epsilon < 0.01$.
All data clearly show stochastic resonance, i.e.\ attain the maximal
signal-to-noise ratio $\SNR_{\mathrm{max}}$ at a noise
strength $D_{\mathrm{max}} > 0$.

\myfig{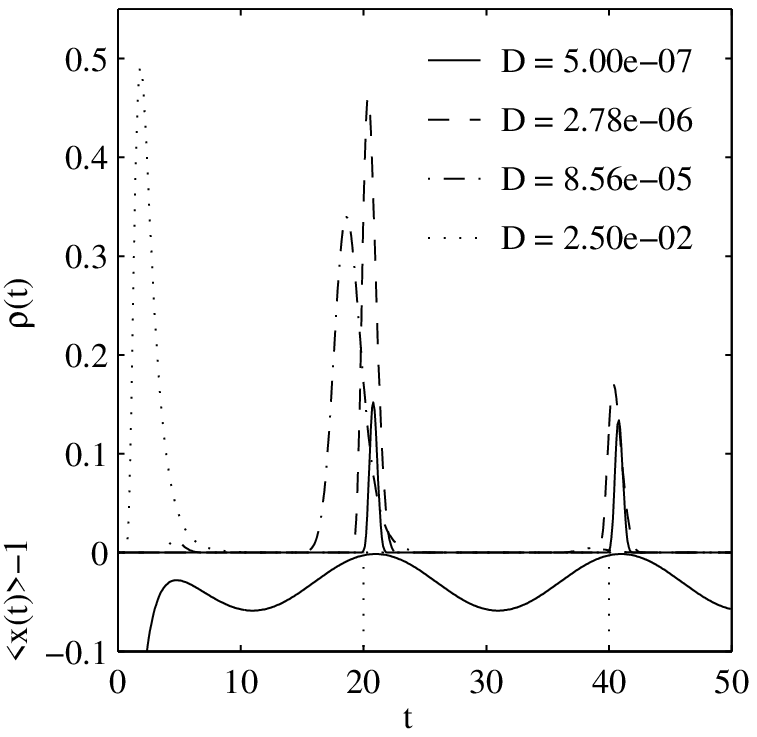}{Deterministic solution and first-passage-time density
for $\mu=0.97$, $q=0.03$, $\omega=0.1\pi$ and $\epsilon=0.0014$ as in
Fig.~\ref{fig_spectra}.  The noise levels correspond to (from top to
bottom): below $D_{\mathrm{max}}$, at $D_{\mathrm{max}}$, at
$D_c$ and at high noise.  The vertical dotted lines mark the first and
second period of the input signal. $\mean{x(t)}$ has been
shifted for clarity.}

To explain why the $\SNR$ peaks, we best turn to the properties of the
deterministic solution $\mean{x(t)}$ and the FPTD $\rho(t)$, which are
shown in Fig.~\ref{fig_fptd} for $\omega = 0.1 \pi$, $\mu = 0.97$,
$q=0.03$ ($\epsilon\approx 0.0014$); this parameter set corresponds to
the solid line in Fig.~\ref{fig_epsvar} and to the spectra shown in
Fig.~\ref{fig_spectra}.  For strong noise, the modulation of
$\mean{x(t)}$ becomes virtually negligible and the threshold crossing
probability is concentrated in a ``drift peak'' at small $t$.  This
drift peak shifts towards $t=0$ and sharpens as the noise strength is
increased, becoming similar to a $\Gamma$-distribution (data not
shown).  In the spectrum, this peak corresponds to a widening hump
shifting towards higher frequencies (Fig.~\ref{fig_spectra}) and no
signal peak is left in the vicinity of the input frequency $\omega$.
As the input noise strength $D$ decreases, threshold crossings become
concentrated around the maxima of $\mean{x(t)}$, and firing events are
synchronized to the input stimulus, with the first peak of $\rho(t)$
dominating the distribution for $D_{\mathrm{max}}$.  As $D$ is reduced
beyond $D_{\mathrm{max}}$, the peak at the first period shrinks and the
firing probability is more evenly distributed over subsequent maxima of
$\mean{x(t)}$.  Therefore, a variable number of maxima is skipped
before the threshold is reached, resulting in erratic firing and thus a
decrease in $\SNR$ (Fig.~\ref{fig_fptd}).

Obviously, we cannot expect stochastic resonance for $\epsilon \le 0$
in this system, for if the deterministic solution $\mean{x(t)}$ reaches
the threshold, spikes will be perfectly synchronized for $D=0$,
although the firing frequency may be far from the frequency of the
input signal.

For small distances-from-threshold ($\epsilon < 0.003$) and low
frequency ($\omega = 0.1\pi$), we observe stochastic resonance at very
small noise strengths $D_{\mathrm{max}}$, and the $\SNR$ decays
algebraically as $D$ is increased beyond $D_{\mathrm{max}}$
(Fig.~\ref{fig_epsvar}).  Furthermore, this decay exhibits a crossover
between two regimes at an intermediate noise strength $D_c$.  For
$D<D_c$, the loss in $\SNR$ is due to the widening of the peaks in the
FPTD, which are located at the maxima of $\mean{x(t)}$, while for
$D>D_c$, the drift peak becomes clearly discernible, corresponding to
the onset of firing not synchronized with the input stimulus, see
Fig.~\ref{fig_fptd}.   

If $\mean{x(t)}$ remains further from thresold, either due to reduced
$q$ or increased $\omega$, stochastic resonance occurs at higher input
noise strengths $D_{\mathrm{max}}$ and yields smaller maximum values
of $\SNR$, see the lower two curves in Fig.~\ref{fig_epsvar}.  This
is to be expected, because as the deterministic solution remains
smaller, the noise contribution to threshold crossing must increase,
reducing the synchronization of firing events with maxima of
$\mean{x(t)}$.

\myfig{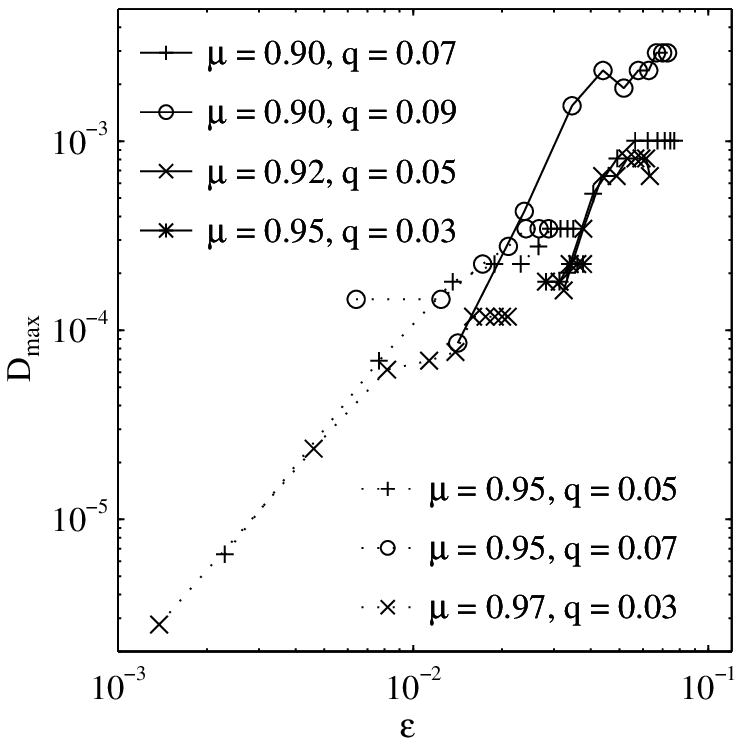}{Position of $\SNR$ maximum vs.\
distance-from-threshold. Data pertaining to identical values of $\mu$
and $q$ but different $\omega$ are connected by lines.  Note that
$D_{\mathrm{max}}$ was chosen from the set of noise strengths for
which calculations were performed, leading to discretization effects
along the ordinate.}

The input noise strength $D_{\mathrm{max}}$ at which $\SNR$ attains
its maximum depends strongly on the distance-from-threshold
$\epsilon$, as is demonstrated in Fig.~\ref{fig_Dmax}.  Here, we have
plotted $D_{\mathrm{max}}$ vs.\ $\epsilon$ on a double-logarithmic
scale.  Indeed, the location of the $\SNR$ maximum roughly obeys a
power law \mbox{$D_{\mathrm{max}} \sim \epsilon^{\gamma}$}.  
A least squares fit yields $\gamma\approx1.5$.  The detailed
dependency of  $D_{\mathrm{max}}$ on $\epsilon$ is quite complex,
though, and not yet well understood.

On the other hand, $D_{\mathrm{max}}$ hardly depends on the input
frequency $\omega$ if the input amplitude $q$ is adjusted so as to
obtain the same distance-from-threshold for all frequencies, see
Fig.~\ref{fig_epsfix}.  This behavior is to be expected from the
mechanism suggested above: the maximal $\SNR$ is reached as the firing
probability is concentrated at the maxima of $\mean{x(t)}$.

%%%%%%%%%%%%%%%%%%%%%%%%%%%%%%%%%%%%%%%%%%%%%%%%%%%%%%%%%%%%%%%%%%%%

\section{Conclusions and Perspectives}
In this letter, we have investigated the response of a model neuron
with reset mechanism to sinusoidal input with additive white noise.
The inter-spike-interval was determined by an efficient numerical
method and power spectral densities were obtained by exploiting the
renewal properties of the spike train generated.  These techniques
permitted us to study the behavior of the model neuron over a wide
range of parameters, especially at very low noise strengths.  We found
clear evidence for stochastic resonance, i.e.\ the signal-to-noise
ratio of the neuron's output shows a distinct maximum at non-vanishing
input noise.  Further, we have proposed a mechanism underlying this
effect.  The results suggest that nature does indeed employ stochastic
resonance to obtain optimal signal-to-noise ratios in an inherently
noisy information processing system.  A detailed comparison with
neurophysiological data will be given elsewhere.

In future work, two questions need to be addressed.  The dependence of
the neuron's response on the phase $\varphi$ of the input stimulus has
yet to be studied in detail.  We expect such work to shed more light
on the detailed structure of the dependencies of the signal-to-noise
ratio on the input noise strength and of the position of the SNR
maximum on the distance-from-threshold.  More importantly, though, our
model shares a weakness with other studies of integrate-and-fire
neurons~\cite{Buls:1996(3958),Buls:1994(4989)}: the presumed phase
reset of the input stimulus is not very plausible from the viewpoint
of neurophysiology.  Work on an extended model overcoming this
difficulty is currently in progress.

%%%%%%%%%%%%%%%%%%%%%%%%%%%%%%%%%%%%%%%%%%%%%%%%%%%%%%%%%%%%%%%%%%%%

\begin{ack}
The authors thank M.~Katakame for inspiring discussions. H.~E.~Plesser
received partial support from Studienstiftung des deutschen Volkes.
\end{ack}

%%%%%%%%%%%%%%%%%%%%%%%%%%%%%%%%%%%%%%%%%%%%%%%%%%%%%%%%%%%%%%%%%%%%

\bibliographystyle{unsrt}
\bibliography{ifneuron_xxx}

\end{document}